%% file: eprint_CIPANP2018.tex
\documentclass[12pt]{article}
\usepackage{graphicx}
\usepackage{amsmath}
\usepackage{epsfig,amssymb,amsfonts,bm}
\usepackage{array}

\newcommand\pubnumber{CIPANP2018-Du}
\newcommand\pubdate{\today}

\newcommand{\bea}{\begin{eqnarray}} 
\newcommand{\eea}{\end{eqnarray}}
\newcommand{\al}{&\!\!\!\!}

\def\napoli{Helmholtz-Institut f\"ur Strahlen- und Kernphysik and Bethe Center for \\Theoretical Physics,
Universit\"at Bonn, Bonn, D-53115, GERMANY}

\def\Title#1{\begin{center} {\Large #1 } \end{center}}
\def\Author#1{\begin{center}{ \sc #1} \end{center}}
\def\Address#1{\begin{center}{ \it #1} \end{center}}

\newcommand\pubblock{\rightline{\begin{tabular}{l} \pubnumber\\
         \pubdate  \end{tabular}}}
\newenvironment{Abstract}{\begin{quotation}  }{\end{quotation}}
\newenvironment{Presented}{\begin{quotation} \begin{center} 
             PRESENTED AT\end{center}\bigskip 
      \begin{center}\begin{large}}{\end{large}\end{center} \end{quotation}}
\def\Acknowledgements{\bigskip  \bigskip \begin{center} \begin{large}
             \bf ACKNOWLEDGEMENTS \end{large}\end{center}}
\input econfmacros.tex
\begin{document}
\begin{titlepage}
\pubblock

\vfill
\Title{Implication of chiral symmetry for the heavy-light meson spectroscopy}
\vfill
\Author{ Meng-Lin Du}
\Address{\napoli}
\vfill
\begin{Abstract}

Many hadronic states observed since 2003, especially for the positive-parity charm-strange states $D_{s0}^\ast (2317)$ and $D_{s1}(2460)$, do not conform with the conventional quark model expectations and raise various puzzles in charm meson spectroscopy. We demonstrate that those puzzles find a natural solution thanks to the recent development of chiral effective theory and Lattice simulations. The existence of the $D_{s0}^\ast (2317)$ and $D_{s1}(2460)$ are attributed to the nonperturbative dynamics of Goldstone bosons scattering off $D$ and $D^\ast$ mesons. It indicates that the lowest positive parity nonstrange scalar charm mesons, the $D_0^\ast(2400)$ in the Review of Particel Physics, should be replaced by two states. The well constructed amplitudes are fully in line with the high quality data on the decays $B^-\to D^+\pi^-\pi^-$ and $D_s^0\to \bar{D}^0K^-\pi^+$. This implies that the lowest positve-parity states are dynamically generated rather than conventional quark-antiquark states. This pattern has also been established for the scalar and axial-vector mesons made from light quarks ($u$, $d$ and $s$ quarks). 

\end{Abstract}
\vfill
\begin{Presented}
The 13th Conference on the Intersections of \\Particle and Nuclear Physics (CIPANP)\\
Palm Springs, CA, USA, May 29--June 3, 2018
\end{Presented}
\vfill
\end{titlepage}
\def\thefootnote{\fnsymbol{footnote}}
\setcounter{footnote}{0}

\section{Introduction}

The hadronic spectrum has received a renewed interest with the recently collected vast amounts of the experimental data, in particular of the exotic states that do not conform with the conventional quark model. Among them the lightest positive parity charm strange mesons $D_{s0}^\ast (2317)$ \cite{Aubert:2003fg} and $D_{s1}(2460)$ \cite{Besson:2003cp} have attracted much attention as they are significantly lighter than their quark model expectations, e.g. 2.48 GeV ad 2.55 GeV, respectively \cite{Godfrey:1985xj,Godfrey:2015dva,Ebert:2009ua}. Moreover, it is observed that the mass difference between the $D_{s0}^\ast(2317)$ and the $D_{s1}(2460)$ is equal to that between the pseudoscalar $D^+$ and the ground state vector $D^{\ast +}$ within 2 MeV. More interestingly, two new charm nonstrange mesons, $D_0^\ast(2400)$ \cite{Abe:2003zm,Link:2003bd} and $D_1(2430)$ \cite{Abe:2003zm} were observed in 2004. Their quantum numbers indicate that they should be the SU(3) partners of the $D_{s0}^\ast (2317)$ and $D_{s1}(2460)$, respectively. However, this assignment immediately raises a puzzle: Why are the masses of the two strange mesons, $D_{s0}^\ast (2317)$ and $D_{s1}(2460)$, almost equal to or even lower than their nonstrange siblings, i.e. the $D_0^\ast(2400)$ and $D_1(2430)$?

Since the discovery of the $D_{s0}^\ast(2317)$ and the $D_{s1}(2460)$, various interpretations of their nature were proposed, including molecules \cite{Barnes:2003dj,
Szczepaniak:2003vy,Guo:2006fu}, tetraquark \cite{Cheng:2003kg,
Chen:2004dy} and chiral partners \cite{Bardeen:2003kt,Nowak:2003ra}. Among them, the hadronic molecular state is of particular interest due to the closeness of their masses to the $DK$ and $D^\ast K$ thresholds, respectively. The low-energy interactions between the charm meson ($D$/$D_s$) and the Goldstone boson $\phi$ ($\pi/K/\eta$) can be systematically studied in the framework of chiral perturbation theory (ChPT), a low-energy effective theory of Quantum Chromodynamics (QCD). The $D_{s0}(2317)$ is interpreted as a $DK$ molecular state by using an unitarized $S$-wave $D$-$\phi$ interaction \cite{Kolomeitsev:2003ac,Guo:2006fu,Gamermann:2006nm}. In those works, only the leading order (LO) amplitudes are employed as the kernels of bubble resummation. Extentions to the next-to-leading order (NLO) were done in Refs.~\cite{Hofmann:2003je,Guo:2008gp,Guo:2009ct,Cleven:2010aw,Guo:2015dha}. 

In the meantime, using the L\"uscher formalism and its extesion to the coupled channels, scattering lengths and phase shifts for the $D\phi$ interaction have been obtained at unphysical quark masses \cite{Liu:2008rza,Liu:2012zya,Mohler:2012na,Mohler:2013rwa,Lang:2014yfa,Moir:2016srx}. The first Lattice calculation only concerns the channels free of disconnected Wick contractions \cite{Liu:2012zya}. In Ref.~\cite{Liu:2012zya}, the obtained scattering lengths are used to determine the low-energy constants (LECs) of the NLO ChPT Lagrangian. In particular, with the LECs determined in Ref.~\cite{Liu:2012zya}, the attraction in channel $(S,I)=(1,0)$, where the $D_{s0}^\ast(2317)$ resides, is strong so that a pole located at $2315_{-28}^{+18}$ MeV emerges \cite{Liu:2012zya}. A recent extension to the $D^\ast$-$\phi$ interaction is performed in Ref.~\cite{Guo:2018gyd} making use of heavy quark spin symmetry. The chiral effective amplitudes were extened to next-to-next-to-leading order (NNLO) in Refs.~\cite{Liu:2009uz,Geng:2010vw,Yao:2015qia,Du:2016ntw,Du:2016xbh,Du:2017ttu,Guo:2018kno}, in which the LECs were determined by $D$ and $D_s$ masses and the lattice results on scattering lengths \cite{Liu:2012zya,Mohler:2012na,Mohler:2013rwa,Lang:2014yfa}. Besides determined by fitting to Lattice data, LECs can also be estimated using the resonance-exchange model, see e.g. in Ref.~\cite{Du:2016tgp}. However, for the coupled-channel on-shell factorization unitarization, the left-hand cut might overlap the unitarity cut and leads to the unitarity violation~\cite{Du:2017ttu}. One way to aviod this problem is to employ the $N/D$ method (see, e.g., Refs.~\cite{Oller:1998zr,Gulmez:2016scm,Du:2018gyn}) to the coupled-channel systems. Due to the overlap of the left- and right-hand cut, the NLO amplitudes in Ref.~\cite{Liu:2012zya} are employed to investigate the possible dynamically generated resonances. In particular, two scalar I=1/2 states are found with the lighter one located more than 100 MeV below its corresponding strange partner \cite{Albaladejo:2016lbb}. In addition, the energy levels computed in Ref.~\cite{Albaladejo:2016lbb} is in a remarkable agreement with the lattice results reported in Ref.~\cite{Moir:2016srx}. It means that the particle listed as $D_0^\ast (2400)$ in the Review of Particle Physics (RPP) \cite{Tanabashi:2018oca} in fact corresponds to two resonances with pole positions $(2105_{-8}^{+6}-i~102_{-12}^{+12})$ MeV and $(2451_{-26}^{+36}-i~134_{-8}^{+7})$ MeV~\cite{Albaladejo:2016lbb}, respectively. In this picture, the puzzle that the experimentally extracted mass of the $D_0^\ast(2400)$ is heavier than its strange partner $D_{s0}^\ast (2317)$ can be easily understood. 

\section{Implication of chiral symmetry for resonances}

One reason why the resonance parameters of the $D_0^\ast(2400)$ and $D_1(2430)$ in the RPP \cite{Tanabashi:2018oca} should be questioned is that the Breit-Wigner (BW) form used to extract them is inconsistent with constraints from the chiral symmetry which requires energy dependent vertices. However, the BW form uses a constant vertex which would lead to a value of the mass larger than its real value. A standard BW parameterization for a $S$-wave $D\pi$ amplitude corresponds to a peak at the invariant energy $E=m_0$ with $m_0$ the BW mass appearing in the BW form if we neglect the mass-dependence of the decay width $\Gamma$ for simplicity. In order to take the chiral constraint into account, we simply modify the parameterization as 
\bea
F^\prime (s) \propto \frac{E_\pi}{s-m_0^2+im_0\Gamma}.
\eea
Then the peak position is shifted from $m_0$ by 
\bea
\Delta =\frac{\Gamma^2 E_D}{4m_0E_\pi-\Gamma^2},
\eea
where the $E_D$ and $E_\pi$ are energy of $D$ and $\pi$ evaluated in the rest-frame of the system. 
For a broad resonance, e.g. $D_0^\ast(2400)$, the shift could be significant. 
However, such a modification can only be applied in a small energy range before the coupled-channel effect becomes important. The broad width of the $D_0^\ast(2400)$ and $D_1(2430)$ require a framework taking both the chiral constraint and the coupled-channel unitarity into account. These requirements are satisfied by the unitarizied ChPT, e.g. in Refs.~\cite{Liu:2012zya,Guo:2015dha}. 

%The effective chiral Lagrangian contributing to $D$-$\phi$ scattering up to NLO can be found in Ref.~\cite{Liu:2012zya}. 
%To describe the possible dynamically generated states, a nonperturbative treatment, i.e. unitarization, is needed. 
The unitarization is equivalent to a resummation of the $s$-channel potentials \cite{Oller:2000fj}
\bea
T^{-1}(s)=V^{-1}(s)+G(s).
\eea
%which is diagrammatically shown in Fig.~\ref{fig:sum}. 
The potential $V(s)$ is derived from the chiral effective Lagrangian and $G(s)$ is the two-point scalar loop functions, regularized with a subtraction constant. The free parameters are determined by fit to lattice data on scattering lengths in 5 disconnected channels \cite{Liu:2012zya}, i.e., $D\pi$ with isospin $I=3/2$, $D\bar{K}$ with $I=0,1$, $D_sK$ and $D_s\pi$. 
%%%%%%%%%%%%%%%%%%%%%%%%%%%%%%%%%%%%%%%%%%%%%%%%%%%%%%%%%%%%%%%%%%%%%%%%%
%\begin{figure}[htb]
%\centering
%\includegraphics[width=1.\textwidth]{uchpt}
%\caption{Diagrammatic illustration of the unitarization.}
%\label{fig:sum}
%\end{figure}
%%%%%%%%%%%%%%%%%%%%%%%%%%%%%%%%%%%%%%%%%%%%%%%%%%%%%%%%%%%%%%%%%%%%%%%%%%%

It was demonstrated that the obtained amplitudes properly predicted the energy levels generated in LQCD \cite{Albaladejo:2016lbb}. This means that the scattering amplitudes for the coupled $D\phi$ system are reasonably based on QCD. Those amplitudes allow us one to identify the poles in the complex energy plane of scattering amplitudes reflecting the possible low-lying positive-parity resonances of QCD in the charm sector, as well as in the bottom secotr when the heavy quark flavor symmetry is employed. The predicted masses for the lowest charm-strange positive-parity mesons are fully in line with the well-estalished measurements, and those for the bottom sectors are consistent with LQCD results, see Table~\ref{tab:ds}. It was also demonstrated that the compositeness (1-Z) \cite{Weinberg:1965zz} for the pole corresponding to the $D_{s0}^\ast(2317)$ related to the $S$-wave scattering length is found to be in the range [0.66,0.73] \cite{Liu:2012zya}. It means that the $D_{s0}^\ast(2317)$ is dominantly a $S$-wave $DK$ molecule. Moreover, employing the heavy quark spin symmetry, $D^\ast K$ interaction is almost same as $DK$ and thus can form a bound state, i.e. $D_{s1}(2460)$, with a similar binding energy \cite{Guo:2009id}
\bea
M_D+M_K-M_{D_{s0}^\ast (2317)}\simeq M_{D^\ast }+M_K-M_{D_{s1}(2460)} %\pm 4~\mathrm{MeV}.
\eea
Moreover, there are two poles, corresponding to two resonances, are found in the channel $(S,I)=(0,1/2)$. The predicted poles, located at the complex energies $\sqrt{s_\text{pole}}=M-i\Gamma/2$, for scalar and axial-vector heavy-light mesons are listed in Table \ref{tab:d}. It is easy to observe that the masses for the lower nonstrange resonances are smaller then those for the strange ones. 

\begin{table}[bt] \caption{Predicted masses of the lowest positive-parity heavy-strange mesons in comparison with the RPP values~\cite{Tanabashi:2018oca} and lattice QCD results (Ref.~\cite{Bali:2017pdv} for charm and Ref.~\cite{Lang:2015hza} for bottom mesons), in units of MeV \cite{Du:2017zvv}.
}
\label{tab:ds}
\vspace{-0.5cm}
\centering
\bea
\begin{array}{cccc} \hline
  \text{meson}    
  & \text{prediction} & \text{RPP} & \text{lattice} \\ \hline
D_{s0}^* & 2315^{+18}_{-28}
& 2317.7\pm 0.6 & 2348^{+7}_{-4} \\
D_{s1} &  2456^{+15}_{-21}  & 2459.5\pm 0.6 &  2451\pm 4
\\
B_{s0}^*   & 5720^{+16}_{-23} & - & 5711\pm 23 \\
B_{s1} &  5772^{+15}_{-21}  & - & 5750\pm 25 \\ \hline
\end{array}
\eea
\end{table}

\begin{table}[t]
\caption{Predicted poles, quoted as ($M,\Gamma/2$), for the positive-parity heavy mesons \cite{Du:2017zvv}. The RPP \cite{Tanabashi:2018oca} values are listed in the last column.
}
\label{tab:d}
\vspace{-0.5cm}
\centering
\bea
\begin{array}{cccc}\hline  
\text{ }    
  & \text{lower pole} & \text{higher pole} & \text{RPP} \\ \hline
D_0^* & \left(2105^{+6}_{-8}, 102^{+10}_{-11}\right) 
& \left(2451^{+35}_{-26},134^{+7}_{-8}\right)  & (2318\pm29,134\pm20)  \\
D_1 &  \left(2247^{+5}_{-6}, 107^{+11}_{-10} \right)  & \left(2555^{+47}_{-30},
203^{+8}_{-9}\right) & (2427\pm40,192^{+65}_{-55})
\\
B_0^*   & \left(5535^{+9}_{-11},113^{+15}_{-17} \right) 
&  \left(5852^{+16}_{-19},36\pm5\right) & - \\
B_1 &   \left( 5584^{+9}_{-11}, 119^{+14}_{-17} \right) 
      &  \left(5912^{+15}_{-18}, 42^{+5}_{-4}\right) & -\\ \hline
\end{array}\nonumber
\eea
\end{table}

\section{Comparison with experimental data}

\begin{figure}
  \bea
  \includegraphics[width=0.31\textwidth]{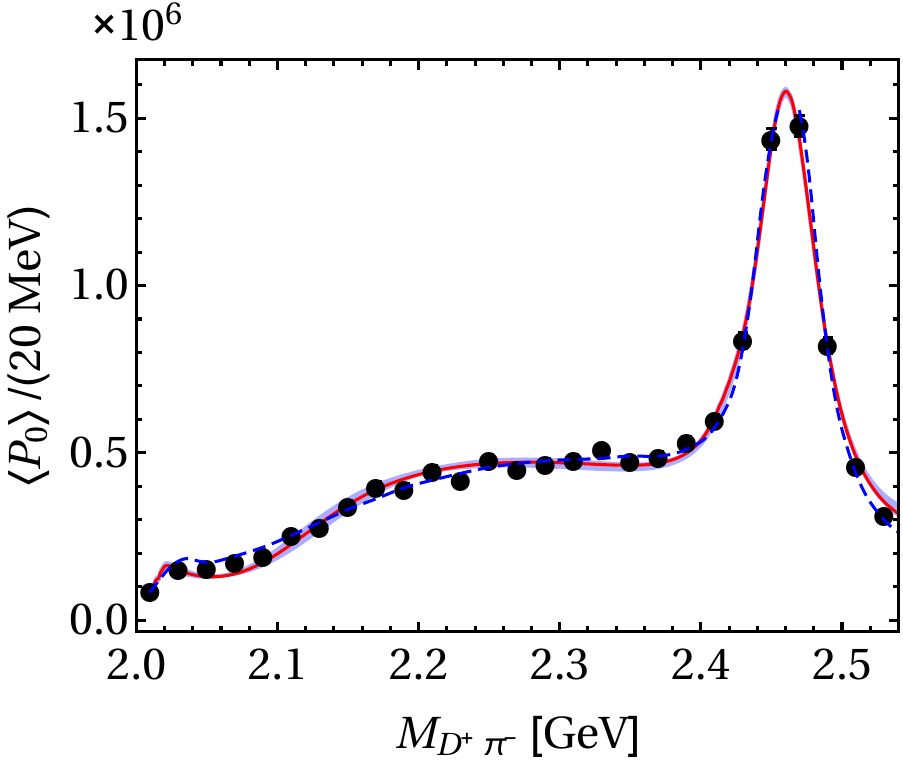} \quad
   \includegraphics[width=0.32\linewidth]{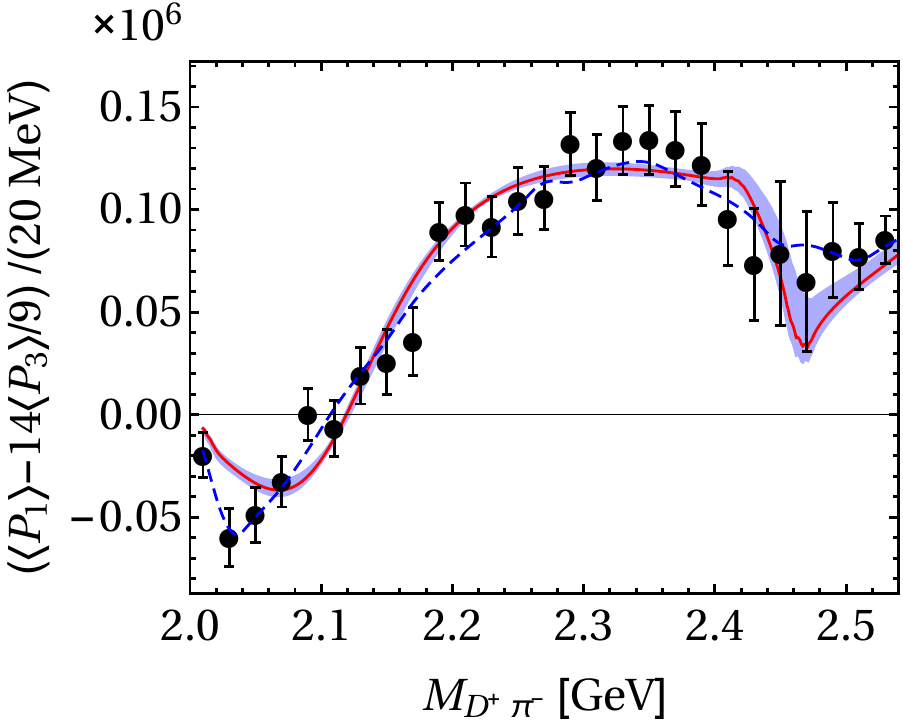}  \quad
  \includegraphics[width=0.31\textwidth]{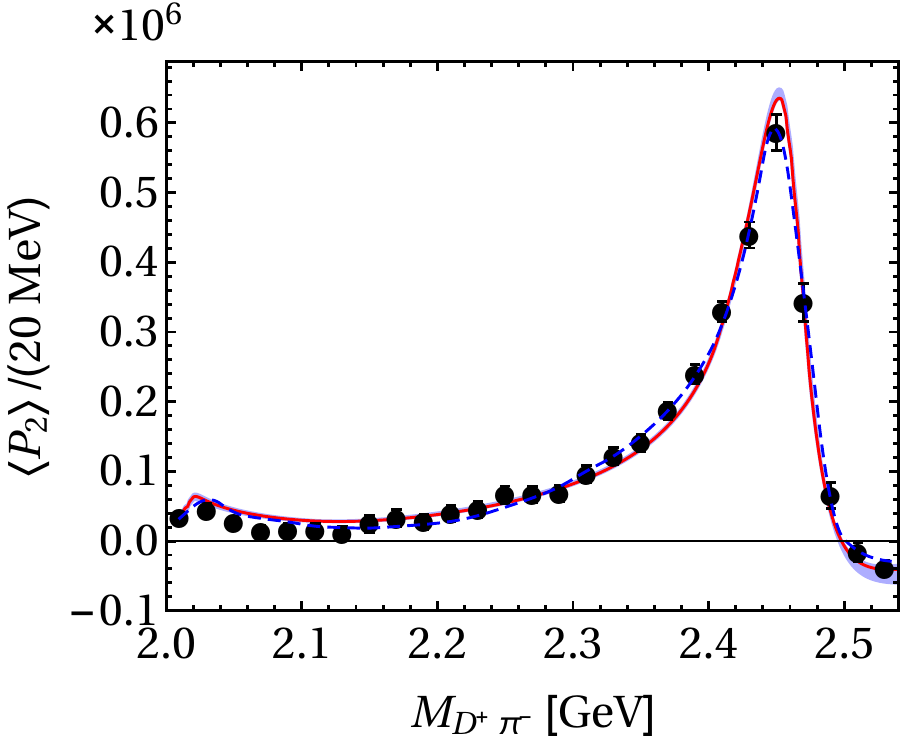}  \nonumber
   \eea
   \caption{Fit to the LHCb data for the angular moments $\langle P_0\rangle$, $\langle P_1\rangle-14\langle P_3\rangle/9$ and $\langle P_2\rangle$ for the $B^-\to D^+\pi^-\pi^-$ reactions \cite{Aaij:2016fma,Du:2017zvv}.}\label{fig:fitD}
\end{figure}

The unitarized chiral $D$-$\phi$ amplitudes are fully in line with the LQCD calculations. They describe the $D_{s0}^\ast (2317)$ and $D_{s1}(2460)$ when the heavy quark symmetry is employed. Moreover, the puzzle of the almost degeneracy of the scalar nonstrange and strange charm mesons is resolved naturally. To show they are fully consistent with the experimental data, we try to describe the reaction $B^-\to D^+\pi^-\pi^-$ \cite{Aaij:2016fma}, which are the best data providing access to the $D\pi$ system at present. When only the low $D\pi$ mass region is concerned, e.g. up to 2.6 GeV, it is sufficient to study the decay amplitude with $S$-, $P$- and $D$-waves, while all channels ($D^+\pi^-$, $D^0\pi^0$, $D^0\eta$ and $D_s^+ K^-$) coupled to $D^+\pi^-$ are considered in the intermediate state. For the $P$- and $D$-wave amplitudes the same BW form as in the LHCb analysis \cite{Aaij:2016fma} are used. For the $S$-wave we employ \cite{Du:2017zvv} 
\bea\label{eq:swave}
	\mathcal{A}_0(s) \al \propto \al  \Big\{ E_\pi \Big[ 2 + G_1(s) \Big( \frac{5}{3} 
	T_{11}^{1/2}(s)+\frac13 T^{3/2}(s) \Big) \Big] + \frac13 E_\eta G_2(s) 
	T_{21}^{1/2}(s)  \nonumber\\ 
	\al \al +\sqrt{\frac23}E_K G_3(s) T_{31}^{1/2}(s) \Big\} 
	+ C E_\eta G_2(s) T_{21}^{1/2}(s), \nonumber
\eea 
where $C=(c_2+c_6)/(c_1+c_4)$ with $c_i$ the low-energy constants in the leading order chiral effective Lagrangian \cite{Du:2017zvv}. Here the $T_{ij}^I(s)$ are the $S$-wave $D\phi$ scattering amplitudes for the coupled-channel system with total isospin $I$, where $i,j$ are channel indices with 1, 2 and 3 referring to $D\pi$, $D\eta$ and $D_s\bar{K}$, respectively. The $G_i(s)$ function is evaluated via a once-subtracted dispersion relation. We fit to the so-called angular moments defined in Ref.~\cite{Aaij:2016fma,Du:2017zvv}, which contain important information about the partial-wave phase variations, up to $M_{D^+\pi^-}=2.54$ GeV for the decay $B^-\to D^+\pi^-\pi^-$ \cite{Aaij:2016fma}. In addition to the $S$-wave amplitude given in Eq.~\eqref{eq:swave}, we include the resonances $D^\ast$ and $D^\ast (2680)$ in the $P$-wave and $D_2(2460)$ in the $D$-wave. Their masses and width are fixed as the central values in the LHCb analysis \cite{Aaij:2016fma}. The best fit has $\chi^2/\text{d.o.f/}=1.7$ and the free parameter values are $C=-3.6\pm0.1$ with the subtraction constant $a_A=1.0\pm 0.1$ for $G_i(s)$. A comparison of the best fit with the LHCb data is shown in the Fig.~\ref{fig:fitD} together with the best fit provided by the LHCb analysis. It is worthwhile to notice that in $\langle P_1\rangle-14\langle P_3\rangle/9$, where the $D_2(2460)$ does not play any role, the data show a significant variation between 2.4 and 2.5 GeV. This feature can be understood as the signal for the opening of the $D^0\eta$ and $D_s^+\pi^-$ thresholds at 2.413 and 2.462 GeV, respectively. Futhermore, the data for the angular moments for $B_s^-\to \bar{D}^0K^-\pi^+$ \cite{Aaij:2014baa} can be easily reproduced in the same freamework \cite{Du:2017zvv}, which has the $\bar{D}_{s0}^\ast(2317)$ as a dynamically generated state.

\section{Summary}

In summary, we have demonstrated that amplitudes fixed from QCD inputs for the $D$-$\phi$ scattering not only resolve the puzzles in charm-meson spectroscopy but also are fully consistent with recent high quality LHCb data on $B$ decays which provide by far the most precise expermental information on the $D\pi$ and $D\bar{K}$ systems. The amplitudes have a pole corresponding to the $D_{s0}^\ast(2317)$ in the $(S,I)=(1,0)$ channel, and two poles in the $(S,I)=(1/2,0)$ channel \cite{Albaladejo:2016lbb}. The lowest charm nonstrange scalar meson, $D_0^\ast(2400)$ listed in the RPP \cite{Tanabashi:2018oca} should be replaced by two poles \cite{Du:2017zvv}. The heavy quark spin symmetry implies that $D_{s1}(2460)$ corresponds to a pole in the $D^\ast K$ scattering amplitude and the $D_1(2430)$ listed in RPP should be replaced by two $J^P=1^+$ states. The coherent picture clearly calls for a change of the paradigm for the positive-parity open-flavor heavy mesons: The lowest positive parity states need to be considered as dynamically generated two-hadron states as opposed to a simple quark-antiquark structure. The similar pattern has been established for the scalar mesons made from light up, down and strange quarks, where the lowest multiplets are considered to be made of states not described by the quark model \cite{Du:2017zvv}.

$~$

\Acknowledgements

The talk is based on the collaboration with Miguel Albaladejo, Pedro Fern\'andez-Soler, Feng-Kun Guo, Christoph Hanhart, Ulf-G. Mei{\ss}ner, Juan Nieves, and De-Liang Yao. Special thanks to Feng-Kun Guo and Ulf-G. Mei{\ss}ner for their reading the manuscript and positive suggestions and comments. This work is partially supported
by the National Natural Science Foundation of China (NSFC) and Deutsche Forschungsgemeinschaft (DFG) through 
funds provided to the Sino--German Collaborative Research Center ``Symmetries and the
Emergence of Structure in QCD'' (NSFC Grant No.~11621131001,
DFG Grant No.~TRR110), by the NSFC (Grant No.~11647601), by the Thousand Talents Plan for Young
Professionals, by the CAS Key Research Program of Frontier Sciences
(Grant No.~QYZDB-SSW-SYS013), by the CAS President's
International Fellowship Initiative (PIFI) (Grant No.~2017VMA0025), by VolkswagenStiftung (Grant No. 93562), by
Spanish Ministerio de Economía y Competitividad
and the European Regional Development Fund
under contracts FIS2014-51948-C2-1-P, FIS2017-84038-C2-1-P and SEV-
2014-0398, by Generalitat Valenciana under contract
PROMETEOII/2014/0068, and by the ``Ayudas para contratos predoctorales para la formación de doctores'' program (BES-2015-072049) from the Spanish MINECO and ESF.

\end{document}

%% file: econfmacros.tex
\def\beq{\begin{equation}}
\def\eeq#1{\label{#1}\end{equation}}
\def\eeqn{\end{equation}}

\def\beqa{\begin{eqnarray}}
\def\eeqa#1{\label{#1}\end{eqnarray}}
\def\eeqan{\end{eqnarray}}

\let\bar=\overbar

\def\Dslash{\not{\hbox{\kern-4pt $D$}}}
\def\dslash{\not{\hbox{\kern-2pt $\del$}}}

\def\msb{{\bar{\ssstyle M \kern -1pt S}}}